\documentstyle[11pt,epsfig]{article}
\begin{document}

\title{{\bf Cosmology with minimal length uncertainty relations  }}
\author{Babak Vakili\thanks{%
email: b-vakili@sbu.ac.ir, bvakili45@gmail.com}\\\\
%EndAName
{\small {\it Department of Physics, Azad University of Chalous,}}\\{\small {\it P. O. Box 46615-397, Chalous, Iran}}}
\maketitle

\begin{abstract}
We study the effects of the existence of a minimal observable
length in the phase space of  classical and quantum de Sitter (dS)
and Anti de Sitter (AdS) cosmology. Since this length has been
suggested in quantum gravity and string theory, its effects in the
early universe might be expected. Adopting the existence of such a
minimum length results in the Generalized Uncertainty Principle
(GUP), which is a deformed Heisenberg algebra between
minisuperspace variables and their momenta operators. We extend
these deformed commutating relations to the corresponding deformed
Poisson algebra in the classical limit. Using the resulting
Poisson and Heisenberg relations, we then construct the classical
and quantum cosmology of dS and Ads models in a canonical
framework. We show that in classical dS cosmology this effect
yields an inflationary universe in which the rate of expansion is
larger than the usual dS universe. Also, for the AdS model it is
shown that GUP might change the oscillatory nature of the
corresponding cosmology. We also study the effects of GUP in
quantized models through approximate analytical solutions of the
Wheeler-DeWitt (WD) equation, in the limit of small scale factor
for the universe, and compare the results with the ordinary
quantum cosmology in each case.\vspace{5mm}\newline PACS numbers:
98.80.-k, 04.60.Ds, 04.60.Kz\vspace{.5cm}\newline Keywords:
Generalized Uncertainty Principle, Noncommutative phase space, Quantum Cosmology
\end{abstract}
\section{Introduction}One of the most important predictions of the
theories which deal with quantum gravity is that there exists a
minimal length below which no other length can be observed
\cite{1}-\cite{6}. From perturbative string theory point of view
{\cite{1,2}}, such a minimal observable length, of order of Planck
scale, is due to the fact that strings cannot probe distances
smaller than the string size. Also, the existence of this minimal
observable length has been suggested in quantum gravity \cite{5},
quantum geometry \cite{7} and black hole physics \cite{8}. Indeed in
the scale of this minimal size {\it i. e.}, in the scales of the
order of Planck length, $l_p=\sqrt{\frac{G\hbar}{c^3}}$, the quantum
effects of gravitation become as important as the electroweak and
strong interactions. Clearly, at low energy levels, these quantum
gravity effects are not too important, but in high energy physics,
that is, the energies of order of Planck mass, $m_p=\hbar/l_p$ such
as very early universe or in the strong gravitational fields of a
black hole, one cannot neglect these effects. In string theory, as
it is indicated in \cite{9}, if the energy of a string reaches the
Planck mass, excitations of the string cause a non-zero extension
and thus it is impossible to measure the position with an
uncertainty smaller than $l_p$, \cite{10}.

One of the interesting features of the existence of a minimum length
described above is the modification it makes to the standard
commutation relation between position and momentum in ordinary
quantum mechanics {\cite{10,11}}, which are called Generalized
Uncertainty Principle (GUP). In one dimension the simplest form of
such relations can be written as
\begin{equation}\label{A}\bigtriangleup p \bigtriangleup x\geq
\frac{\hbar}{2}\left(1+\beta (\bigtriangleup
p)^2+\gamma\right),\end{equation} where $\beta$ and $\gamma$ are
positive and independent of $\bigtriangleup x$ and $\bigtriangleup
p$, but may in general depend on the expectation values $<x>$ and
$<p>$. The usual Heisenberg commutation relation can be recovered in
the limit $\beta=\gamma=0$. As is clear from equation (\ref{A}),
this equation implies a minimum position uncertainty of
$(\bigtriangleup x)_{min}=\hbar \sqrt{\beta}$, and hence $\beta$
must be related to the Planck length. For a more general discussion
of such deformed Heisenberg algebras, especially in three dimension,
see Ref. \cite{12}. Now, it is possible to realize equation
(\ref{A}) from the following commutation relation between position
and momentum operators
\begin{equation}\label{B} \left[x,p\right]=i\hbar \left(1+\beta
p^2\right),\end{equation} where we take $\gamma=\beta <p>^2$. More
general cases of such commutation relations are studied in Refs.
\cite{10}-\cite{14}.

The low energy effects of the modified Heisenberg uncertainty
relations have been extensively  studied in various phenomena in
recent years. For example, the spectrum of Hydrogen atom arisings
from modified Schr\"{o}dinger equation based on GUP is studied in
\cite{15} and \cite{16}. In \cite{15}, using this approach for the
Hydrogen atom results in the splitting of degenerate energy levels.
Also, an upper bound for the deformation parameter of Heisenberg
algebra is obtained. On the other hand, the generalized
Schr\"{o}dinger equation and its momentum space representation have
been studied in \cite{16} and \cite{17}. In this direction, in
\cite{18}, the authors consider two well-known problems in usual
quantum mechanics, a particle restricted to move in a one
dimensional box and the free particle, and solved them in the GUP
framework. In each case the resulting wave functions from the
corresponding modified Schr\"{o}dinger equation are compared with
the ordinary solutions. Applying this formalism to quantum
fluctuations in the early universe and the corresponding effects on
the inflation is investigated in \cite{19}.

It is a generally accepted practice to introduce GUP either through
the coordinates or fields which may be called geometrical or phase
space GUP respectively \cite{Ha}. Applying GUP to ordinary quantum
field theories where geometry is considered to obey the GUP
relations are interesting to to study since they could provide an
effective theory bridging the gap between ordinary quantum field
theory and string theory, currently considered as the most important
choice for quantization of gravity. A different approach to GUP is
through its introduction in the phase space constructed by
minisuperspace fields and their conjugate momenta \cite{Ba}. Since
cosmology can test physics at energies that are much higher than
those on Earth, it seems natural that the effects of quantum gravity
could be observed in this context. On the other hand in cosmological
systems, since the scale factor, matter fields and their conjugate
momenta play the role of dynamical variables of the system,
introducing GUP in the corresponding phase space is particularly
relevant.

In this paper, we study the well-known classical and quantum de
Sitter (dS) and Anti de Sitter (AdS) cosmologies in a canonical
framework within the context of GUP in phase space. Here we extend
the deformed Heisenberg algebra to the classical limit to get a
deformed Poisson algebra, that is, the classical version of GUP. We
then resolve the classical dS and AdS cosmologies by using this
deformed Poisson brackets. The resulting solutions show some
differences between the corresponding cosmologies and the usual ones
depending on the deformation parameter $\beta$. Also, we consider
the quantum cosmology of these models and show that applying GUP to
the corresponding phase space results in a fourth order
Wheeler-DeWitt (WD) equation for the corresponding quantum
cosmology. Although, we cannot solve this fourth order differential
equation in general, we can provide some approximate analytical
solutions of this equation in the limit of small scale factors.
Finally, we compare the wave functions of ordinary dS and AdS
universes with the wave functions obtained from applying the GUP in
this context. It is to be noted that our presentation does not claim
to clear the role of GUP in cosmology in a fundamental way because
we just study the problem in a special simple model. However, this
may reflect realistic scenarios in similar investigations which deal
with this problem in a more fundamental way. In what follows, we
work in units where $c=\hbar=16 \pi G=1$.
\section{Classical dS and AdS Cosmologies with Deformed Poisson
Algebra} To start, let us make a quick review of the well-known dS
and AdS cosmologies and obtain the mini-superspace Lagrangian and
Hamiltonian. We assume that the universe is homogeneous and
isotropic. Thus, it can be described by the flat Robertson-Walker
metric
\begin{equation}\label{C}
ds^2=-dt^2+a^2(t)(dx^2+dy^2+dz^2),\end{equation}
where $a(t)$ is
the scale factor of the universe. The scalar curvature
corresponding to metric (\ref{C}) is
\begin{equation}\label{D} {\cal
R}=6\left(\frac{\ddot{a}}{a}+\frac{\dot{a}^2}{a^2}\right),
\end{equation}
where a dot represents differentiation with respect to cosmic time
$t$. To construct the canonical formalism of the theory, let us
start with the Einstein-Hilbert action
\begin{equation}\label{E} {\cal S}=\int ({\cal R}
-2\Lambda)\sqrt{-g}d^4x,
\end{equation}
where $g$ is the determinant
of the spacetime metric and $\Lambda$ is the cosmological constant
representing the vacuum energy. Substituting (\ref{C}) and
(\ref{D}) into (\ref{E}) and integrating over the spacial
dimension, we are led to the following Lagrangian
\begin{equation}\label{F} {\cal L}=a\dot{a}^2+\frac{1}{3}\Lambda
a^3,
\end{equation}
which yields the Hamiltonian
\begin{equation}\label{G}
{\cal H}=\frac{p_a^2}{4a}-\frac{1}{3}\Lambda a^3,
\end{equation}where $p_a$ is the momentum conjugate to $a(t)$.
The simplest classical inflationary and oscillatory models, say dS
and AdS models respectively, can be obtained in this step using the
above Hamiltonian to construct the equations of motion, but a remark
about this Hamiltonian is the factor ordering problem when one
embarks on construction a quantum mechanical operator equation. This
is an indication that in quantizing the system, the ordering problem
becomes important. In ordinary canonical quantum cosmology one can
deal with a parameter denoting the ambiguity in the ordering of
factors to guarantee Hermiticity of the operator corresponding to
the Hamiltonian. But as we shall see in the next section,
quantization of the model in the GUP framework needs a more
complicated representation for momentum operator and thus
introducing the factor ordering parameter in this representation is
not an easy task. Therefore, to transform Hamiltonian (\ref{G}) to a
more manageable form, consider the change of variables
$(a,p_a)\rightarrow (u,p_u)$ as \footnote{note that at $a=0$ we have
also $p_a=0$ and thus $p_u$ is well defined at $a=0$ {\it i. e.} at
$u=0$.}
\begin{equation}\label{H}
u=a^{3/2}\Rightarrow p_u=\frac{2}{3}a^{-1/2}p_a.
\end{equation}
In terms of these new
variables the Hamiltonian takes the form
\begin{equation}\label{I}
{\cal H}=\frac{9}{16}p_u^2-\frac{1}{3}\Lambda
u^2.
\end{equation}
Now, use of the usual equation of motion $\dot{X}=\{X,H\}$, and that
of the canonical variables satisfying the Poisson algebra
\begin{equation}\label{J}
\{x_i,p_j\}=\delta_{ij},\hspace{.5cm}\{x_i,x_j\}=\{p_i,p_j\}=0,
\end{equation}
one obtains the well-known inflationary dS solution in the case of
a positive cosmological constant
\begin{equation}\label{K} u(t)=e^{\Omega t},
\end{equation}
in which $\Omega^2=\frac{3}{4}\Lambda$ and we take $u(t=0)=1$.
Now, we would like to investigate the effects of classical version
of GUP, {\it i.e.}, classical version of commutation relation
(\ref{B}) on the above cosmology. As is well known, in the
classical limit the quantum mechanical commutators should be
replaced by the classical Poisson brackets as $[P,Q]\rightarrow
i\hbar \{P,Q\}$. Thus, considering the GUP issue in classical
phase space changes the Poisson algebra (\ref{J}) into their
deformed forms as
\begin{equation}\label{L}
\left\{x_i,x_j\right\}=\left\{p_i,p_j\right\}=0,\hspace{.5cm}
\left\{x_i,p_j\right\}=\delta_{ij}\left(1+\beta
p^2\right).
\end{equation}Such deformed Poisson algebra is used in \cite{Ben} to
investigate effects of the deformation on the classical orbits of
particles in a central force field and on the Kepler third law.
Also, the stability of planetary circular orbits in the framework of
such deformed Poisson brackets is considered in \cite{Ku}. Note that
here we deal with modifications of a classical cosmology that become
important only at the Planck scale, where the classical description
is no longer appropriate and a quantum model is required. However,
before quantizing the model we shall provide a deformed classical
cosmology. In this classical description of the universe in
transition from commutation relation (\ref{B}) to Poisson brackets
(\ref{L}) we keep the parameter $\beta$ fix as $\hbar \rightarrow
0$. In string theory this means that the string momentum scale is
fixed when its length scale approaches the zero. Therefore, by
extending the above Poisson algebra to the phase space constructed
by $u$ and $p_u$, we should obtain the corresponding cosmology
resulting from Hamiltonian (\ref{I}) where its canonical variables
$u$ and $p_u$ satisfy the Poisson bracket
\begin{equation}\label{M}
\left\{u,p_u\right\}=1+\beta p_u^2,
\end{equation}
which yields the following equations of motion
\begin{equation}\label{N}
\dot{u}=\left\{u,{\cal H}\right\}=\frac{9}{8}p_u\left(1+\beta
p_u^2\right),\end{equation}
\begin{equation}\label{O}
\dot{p_u}=\left\{p_u,{\cal H}\right\}=\frac{2}{3}\Lambda u
\left(1+\beta p_u^2\right).
\end{equation}
Differentiation of equation (\ref{N}) and use of equation (\ref{O})
yields
\begin{equation}\label{P}
\ddot{u}=\frac{3}{4}\Lambda u \left(1+4\beta p_u^2\right),
\end{equation}
where we consider only the terms of first order in $\beta$. On the
other hand from the system of differential equation (\ref{N}) and
(\ref{O}) we obtain
\begin{equation}\label{R}
\frac{\dot{u}}{\dot{p_u}}=\frac{27}{16\Lambda}\frac{p_u}{u},
\end{equation}
which can be immediately integrated with the result
\begin{equation}\label{S}
p_u^2=\frac{16}{27}\Lambda u^2,
\end{equation}
where we take the integration constant to be zero. Now, substituting
the result (\ref{S}) into equation (\ref{P}), we are led to the
following differential equation for $u(t)$
\begin{equation}\label{T}
\ddot{u}-\frac{3}{4}\Lambda u -\frac{16}{9}\beta \Lambda^2
u^3=0.
\end{equation}This is the deformed version of equation of motion and the
last term is the GUP correction, had it not been for this term, we
would have got a pure dS solution as (\ref{K}). The nonlinear term
in the above equation, because of the presence of the deformation
parameter $\beta$, is very small and thus on this assumption we may
solve the equation approximately by a perturbation method. The
classical solution (\ref{K}) with a good approximation is equal to
the solution of the above equation, hence, to first order of
approximation we can substitute solution (\ref{K}) into the
$\beta$-term to obtain the following equation
\begin{equation}\label{U}
\ddot{u}-\Omega^2 u=\left(\frac{16}{9}\right)^2\beta \Omega^4
e^{3\Omega t},
\end{equation}
where as before $\Omega^2=\frac{3}{4}\Lambda$, $\Lambda>0$. Thus, we
see that in our linearization mechanism, the effect of the deformed
Poisson algebra is to change the equation of motion to a
nonhomogeneous differential equation. The complete solution of
equation (\ref{U}) can be easily obtained by adding its particular
solution to the homogeneous solution (\ref{K}), with the result
\begin{equation}\label{V}
u(t)=e^{\Omega t}+\frac{32}{81}\beta \Omega^4 e^{3\Omega
t}.
\end{equation}
Figure 1 shows the effect of $\beta$ on the expansion rate of the
universe. As it is clear from this figure, increasing the value of
$\beta$ results in larger expansion rate and thus larger size for
the universe compared to the case when $\beta=0$, at equal times.
This phenomenon may be interesting in the inflationary cosmological
scenarios.
\begin{figure}\begin{center}
\epsfig{figure=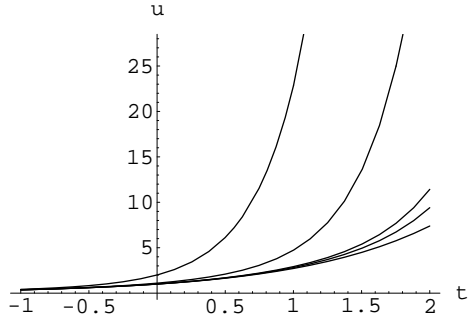,width=7cm} \caption{\footnotesize The effect
of $\beta$ on the expansion rate of the universe for a positive
cosmological constant. We take the numerical values $\Omega =1$ and
$\frac{32}{81}\beta=0, .005, .01, .1, 1$ from bottom to top
respectively. } \label{fig1}\end{center}
\end{figure}
Now, let us deal with the case of a negative cosmological constant,
{\it i. e.}, the AdS model. In this case the solutions can be
obtained with the replacement
$\Omega^2\rightarrow-\omega^2=-\frac{3}{4}|\Lambda|$. Thus, the
ordinary cosmology is an oscillatory universe with
\begin{equation}\label{W}
u(t)=\sin \omega t,
\end{equation}
where we have taken the initial condition $u(0)=0$. Substituting
this solution into the negative cosmological constant version of
(\ref{U}), that is,
\begin{equation}\label{X}
\ddot{u}+\omega^2 u=\left(\frac{16}{9}\right)^2\beta \omega^4
\sin^3 \omega t,
\end{equation}
we are led to the following solution
\begin{equation}\label{Y}
u(t)=\sin \omega t-\frac{32}{27}\beta \omega^3 t \cos \omega
t+\frac{8}{81}\beta \omega^2 \sin 3\omega t.
\end{equation}
We show the corresponding scale factors for some values of $\beta$
in figure 2. As this figure shows, increasing the value of $\beta$
may disturb the oscillatory nature of the universe. Also, in the
presence of $\beta$ terms, the period of oscillations become
larger and thus the Big-Crunch in the corresponding cosmological
model occurs later in comparison with the ordinary AdS model in
which $\beta=0$.
\begin{figure}\begin{center}
\epsfig{figure=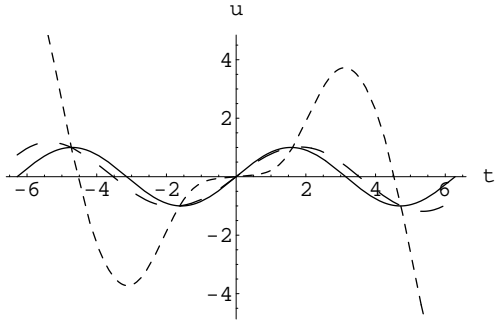,width=7cm} \caption{\footnotesize The
effect of $\beta$ on the scale factor of the universe for a
negative cosmological constant. The figure is plotted for
numerical values $\omega=1$ and $\beta=0, .1, 1$ for solid, large
dashed and small dashed curves respectively.}
\label{fig2}\end{center}
\end{figure}

\section{Quantum dS and AdS Cosmologies with GUP}
In general GUP in its original form (see \cite{10,11}) implies a
noncommutative underlying geometry for space time. But formulation
of gravity in a noncommutative space time is highly nonlinear and
setting up cosmological models is not an easy task. Here our aim
is to study some aspects regarding the application of the GUP
framework in quantum cosmology, {\it i. e.} in the context of a
minisuperspace reduction of the dynamics. As is well-known in the
minisuperspace approach of quantum cosmology, which is based on
the canonical quantization procedure, one first freezes a large
number of degrees of freedom by imposition of symmetries on the
spacial part of the metric and then quantizes the remaining ones.
Therefore, in the absence of a full theory of quantum gravity,
quantum cosmology is a quantum mechanical toy model with a finite
degrees of freedom which is a simple arena to test ideas and
constructions which can be introduced in quantum general
relativity. In this respect, the GUP approach to quantum cosmology
appears to have physical grounds. In fact, one notes that a
deformation of the canonical Heisenberg algebra immediately leads
to a generalized uncertainty principle. In other words, the GUP
scheme relies on a modification of the canonical quantization
prescriptions and, in this respect, it can be reliably applied to
any dynamical system (see \cite{Mon} for a more clear explanation
on the GUP in the minisuperspace dynamics). In this sense, as we
have mentioned in introduction there are various realizations of
GUP and one can introduce deformation between different dynamical
variables of the corresponding minisuperspace and of course get
different results. But we use the most common approaches which are
used in the literature.

We now focus attention on the study of the quantum cosmology of the
models described above. For this purpose we quantize the dynamical
variables of the model with the use of the WD equation, that is,
${\cal H}\Psi=0$, where ${\cal H}$ is the operator form of the
Hamiltonian given by equation (\ref{I}), and $\Psi$ is the wave
function of the universe, a function of the scale factors and the
matter fields, if they exist. In ordinary canonical quantum
cosmology, use of usual commutation relation $[u,p_u]=i$, results in
the well-known representation $p_u=-i
\partial/\partial u$, to construct the WD equation. But in the
GUP framework, as we have mentioned in the introduction, the
existence of a minimum observable length requires the following
commutation relation between $u$ and $p_u$,
\begin{equation}\label{Z}
\left[u,p_u\right]=i\left(1+\beta p_u^2\right).
\end{equation}
Of course, this new algebra between canonical variables changes
the representations of the corresponding operators. Indeed, in the
GUP formalism the following representation of the momentum
operator in $u$- space fulfills the relation (\ref{Z}) up to first
order in $\beta$, \cite{17}
\begin{equation}\label{AB}
p_u=-i\left(1-\frac{\beta}{3}\frac{\partial^2}{\partial
u^2}\right)\frac{\partial}{\partial u}.
\end{equation}
Now, using this representation for the momentum operator, the WD
equation can be written up to first order in $\beta$ as
\begin{equation}\label{AC}
\left[\frac{2}{3}\beta \frac{d^4
}{du^4}-\frac{d^2}{du^2}-\frac{16}{27}\Lambda u^2\right]\Psi(u)
=0.
\end{equation}
The appearance of a differential equation of fourth order to
describe a physical phenomenon is interesting, since it requires to
investigate the corresponding new boundary conditions, which is not
the goal of our study in this paper. Instead, since we cannot solve
the above equation analytically, we provide an approximate method
which in its validity domain we need to solve only a second order
differential equation. Before trying  this, we should give some
comments about the quantum mechanical wave functions in GUP
formalism. As it is clearly discussed in \cite{10}, in the GUP
framework, because of the existence of a minimal observable length,
we cannot have localized quantum states. In this formalism one
resolves this problem by considering maximal localization which are
proper quasi-position wave functions $\Psi(u)$. Indeed, these states
can be used to define a quasi-position representation, which has a
direct interpretation in terms of position measurements. For a more
detailed description of this important issue see
{\cite{10,15,16,17}}.

Taking $\beta=0$ in equation (\ref{AC}) yields the ordinary WD
equation where its solutions, in the case of a positive
cosmological constant, can be written in terms of Bessel functions
as
\begin{equation}\label{AD}
\Psi(u)=u^{1/2}\left[c_1J_{1/4}\left(\frac{\omega}{2}u^2\right)+
c_2Y_{1/4}\left(\frac{\omega}{2}u^2\right)\right],
\end{equation}
where $\omega^2=\frac{16}{27}\Lambda$, $\Lambda>0$. To avoid
divergent behavior of the wave function in the limit $u\rightarrow
0$, we take $c_2=0$, \cite{20}. Therefore, a suitable wave function
in this case reads
\begin{equation}\label{AE}
\Psi(u)=u^{1/2}J_{1/4}\left(\frac{\omega}{2}u^2\right).
\end{equation}
In the case when $\beta \neq 0$, as we mentioned before, equation
(\ref{AC}) cannot be solved exactly, but if we use the solution
(\ref{AE}) in the $\beta$-term of (\ref{AC}), we may obtain some
approximate analytical solutions in the region $u\rightarrow 0$. To
this end, note that the effects of $\beta$ are important at Planck
scales, {\it i. e.} in cosmology language in the very early
universe, that is, when the scale factor is small $u\sim 0$. The
limiting behavior of solution (\ref{AE}) in the region $u\sim 0$ is
\cite{20}
\begin{eqnarray}\label{AF}
\Psi(u)=u^{1/2}J_{1/4}\left(\frac{\omega}{2}u^2\right)\rightarrow\nonumber \\
u^{1/2}\left\{\frac{1}{\Gamma(5/4)}
\left(\frac{\omega}{4}u^2\right)^{1/4}-\frac{1}{\Gamma(9/4)}
\left(\frac{\omega}{4}u^2\right)^{1/4+2}+...\right\} &=&\nonumber \\
\\
u^{1/2}\left\{\left(\frac{\omega}{4}\right)^{1/4}\frac{1}{\Gamma
(5/4)}u^{1/2}-\left(\frac{\omega}{4}\right)^{9/4}\frac{1}{\Gamma(9/4)}u^{9/2}\right\}.\nonumber
\end{eqnarray}
Thus, in the limit $u\rightarrow 0$, we have a wave function of the
form
\begin{equation}\label{AG}
\Psi(u)=C_1u-C_2u^5,
\end{equation}
and therefore its fourth derivative is equal to
$d^4\Psi/du^4=-120C_2u$. Now, if we consider only the first order
term with respect to $u$ in (\ref{AG}), we get $d^4\Psi/d^4
u=-6\omega^2 \Psi.$ Substituting this result into equation
(\ref{AC}) leads us to the following WD equation for the
corresponding dS quantum cosmology in the GUP framework
\begin{equation}\label{AH}
\frac{d^2\Psi}{du^2}+\omega^2 u^2 \Psi+4\beta \omega^2
\Psi=0.
\end{equation}
This equation, after a change of variable $v=i\omega u^2$ and
transformation $\Psi=v^{-1/4}\phi$, takes the form
\begin{equation}\label{AI}
\frac{d^2\phi}{dv^2}+\left(-\frac{1}{4}+\frac{\kappa}{v}+\frac{1/4-\mu^2}{v^2}\right)\phi=0,
\end{equation}
where $\kappa=-i\beta \omega$ and $\mu=1/4$. The above equation is
the well-known Whittaker differential equation and its solutions
can be written in terms of Confluent Hypergeometric functions
$M(a,b;x)$ and $U(a,b;x)$ as
\begin{equation}\label{AJ}
\phi(v)=e^{-v/2}v^{\mu
+1/2}\left\{c_1M\left(\mu-\kappa+\frac{1}{2},2\mu+1;v\right)+
c_2U\left(\mu-\kappa+\frac{1}{2},2\mu+1;v\right)\right\},
\end{equation}
Therefore the solutions of equation (\ref{AH}) read
\begin{equation}\label{AL}
\Psi(u)=e^{-i\omega u^2/2}\left(i\omega
u^2\right)^{1/2}\left\{c_1M\left(\frac{3}{4}+i\beta
\omega,\frac{3}{2};i\omega u^2\right)+c_2U\left(\frac{3}{4}+i\beta
\omega,\frac{3}{2};i\omega u^2\right)\right\}.
\end{equation}
Bearing in the mind that if $\beta=0$, the solution (\ref{AE})
should be recovered, we take $c_2=0$ and retain only the function
$M(a,b;x)$ \cite{20}. Therefore, we are led to the final form of the
GUP wave function of the early universe as follows
\begin{equation}\label{AM}
\Psi(u)=e^{-i\omega u^2/2}\left(i\omega
u^2\right)^{1/2}M\left(\frac{3}{4}+i\beta \omega,\frac{3}{2};i\omega
u^2\right).
\end{equation}
In figure 3 we have plotted the square of wave functions (\ref{AE})
and (\ref{AM}), which are related to the cases $\beta=0$ and $\beta
\neq 0$, respectively. As is clear from this figure, in both cases
the probability (which is proportional to $|\Psi(u)|^2$), of the
emerging universe from $u=0$ is zero, which is in agreement with
classical solutions. Indeed the peaks in $|\Psi(u)|^2$ occur after
$u=0$ which shows an expanding universe as the classical solutions
also predict. But as the figure shows, the square of the wave
function in the case of $\beta \neq 0$ has a more rapid slope in
comparison with the case $\beta=0$. We recall the same behavior for
the classical solutions for which the $\beta \neq 0$ solutions have
a larger rate of expansion.
\begin{figure}\begin{center}
\epsfig{figure=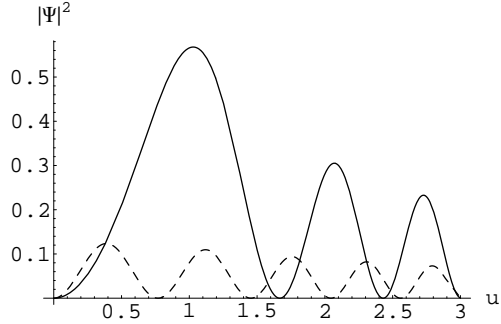,width=7cm} \caption{\footnotesize The
square of wave function of the dS quantum universe. We take the
numerical values $\omega=2$ and $\beta=0,1$ for solid and dashed
curves respectively.} \label{fig3}\end{center}
\end{figure}

The solutions of the WD equation in the case of $\Lambda<0$, {\it i.
e.} the AdS quantum cosmology wave functions can be obtained by
replacing $\omega \rightarrow i\omega$. Thus, when $\beta=0$ the
wave function of the WD equation can be written in terms of modified
Bessel functions as
\begin{equation}\label{AN}
\Psi(u)=u^{1/2}\left\{c_1K_{1/4}\left(\frac{\omega}{2}u^2\right)+
c_2I_{1/4}\left(\frac{\omega}{2}u^2\right)\right\}.
\end{equation}
The functions $I_{\nu}(x)$ are usually omitted because of their
divergent behavior in the limit $u \rightarrow \infty$. Therefor,
the wave function in this case reads
\begin{equation}\label{AO}
\Psi(u)=u^{1/2}K_{1/4}\left(\frac{\omega}{2}u^2\right).
\end{equation}
Now, with the GUP consideration in this quantum cosmology model,
{\it i. e.}, taking $\beta \neq 0$ and following the same procedure
as in the positive cosmological constant case, we obtain the
corresponding WD equation in the limit $u \sim 0$, as
\begin{equation}\label{AP}
\frac{d^2\Psi}{du^2}-\omega^2 u^2 \Psi+4\beta
\omega^2\Psi=0,
\end{equation}
where its solutions can be again written in terms of the Confluent
Hypergeometric functions as
\begin{equation}\label{AR}
\Psi(u)=e^{-\omega u^2/2}\left(\omega
u^2\right)^{1/2}U\left(\frac{3}{4}-\beta \omega,\frac{3}{2};\omega
u^2\right).
\end{equation}
However this time to recover the solution (\ref{AO}) in the limit
$\beta=0$, we retain only $U(a,b;x)$ function \cite{20}. Figure 4
shows $|\Psi(u)|^2$ for typical values of parameters in both
$\beta=0$ and $\beta \neq 0$ cases. We see that in ordinary AdS
quantum cosmology, the universe with highest probability emerges at
$u=0$. This behavior also occurs in the model with GUP, but in spite
of taking $\beta=0$, in this case the square of the wave function
has several peaks. The emergence of new peaks in the GUP model wave
function may be interpreted as a representation of different quantum
states that may communicate with each other through tunnelling. This
means that there are different possible states from which our
present universe could have evolved and tunnelled in the past.
Similar behavior also occurs in noncommutative cosmological
scenarios \cite{21}, which may mean that there is a close
relationship between noncommutativity and GUP \cite{22}.
\begin{figure}\begin{center}
\epsfig{figure=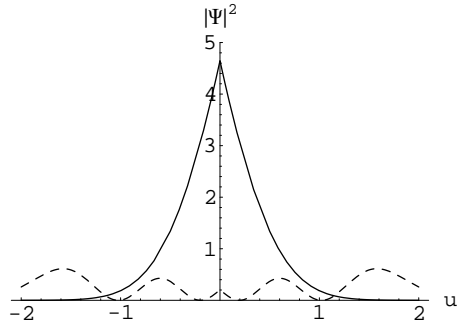,width=7cm} \caption{\footnotesize The
square of the wave function for AdS quantum universe. We take the
numerical values $\omega=2$ and $\beta=0,1$ for solid and dashed
curves respectively.} \label{fig}\end{center}
\end{figure}
\section{Conclusions}
In this paper we have studied the effects of generalized uncertainty
relation on phase space, in classical and quantum dS and AdS
cosmologies. This yields a deformed Heisenberg commutation relation
between the phase space operators called GUP. We extend this new
algebra to the classical limit to get the corresponding Poisson
brackets. With this classical version of GUP, we obtain the
corrections to dS and AdS cosmologies. We have shown that if the
deformation parameter $\beta$ is not equal to zero, the resulting dS
cosmology exhibits an inflationary universe in which the rate of
expansion of the universe is larger than the ordinary dS model.
Also, in the classical AdS model there is some differences between
ordinary and deformed models. In the early universe limit
($t\rightarrow 0$), both models have oscillatory behavior but the
model based on the deformed Poisson algebra has a larger period,
which means that in such models the corresponding Big-Crunch occur
later in comparison to the ordinary model. The oscillatory behavior
of the classical AdS model can be disturbed in the presence of a
$\beta \neq 0$, when time increases. We have also studied the
corresponding quantum cosmologies through finding the exact
solutions of the WD equation in ordinary dS and AdS models and its
approximate analytical solutions in the case when GUP is considered.
We have seen that in the quantum dS model the square of wave
function has a more rapid slope if the model has a nonzero
deformation parameter which may be interpreted as a universe with
larger rate of expansion, in agreement with the classical model.
Finally, in the quantum AdS model we have shown that in the presence
of GUP, the square of wave function of the universe has several
peaks which may be related to different states in the early universe
from which our present universe could have evolved. The ordinary AdS
models do not exhibit such behavior.
\vspace{5mm}\newline \noindent {\bf
Acknowledgement}\vspace{2mm}\noindent\newline The author would like to thank the research council of
Azad University of Chalous for financial support.

\end{document}